\documentclass[acmsmall,nonacm]{acmart}
\usepackage{xspace}

\usepackage{caption}
\usepackage{subcaption}
\usepackage{enumitem}

\usepackage{tikz}
\usepackage{tikz-cd}
\usetikzlibrary{tqft}
\usetikzlibrary{shapes.geometric,tqft}

\newcommand{\cohop}{{\sc CoHOp}\xspace}
\newcommand{\dysco}{{\sc DySCo}\xspace}

\theoremstyle{definition}
      \newtheorem{definition}{Definition}
            \newtheorem{lemma}[definition]{Lemma}

\theoremstyle{plain}
      \newtheorem{theorem}[definition]{Theorem}

\theoremstyle{remark}

\usepackage{todonotes}

\usepackage{macros-acm}

\def\F{\mathcal{F}}
\def\Fb{\overline{\F}}

\newcommand{\project}{\cat[Project]}

\newcommand{\grph}{\cat[Grph]}
\newcommand{\igrph}{\intC{\grph}}

\newcommand{\quotient}[2]{(#1)_{/#2}}
\newcommand{\set}[1]{\{#1\}}

\def\I{[0;1]}

\title{Linear Realizability and Cobordisms}

\author{Valentin Maestracci}
\orcid{0000-0003-0037-9041}

\affiliation{
  \department[0]{Mathematics (I2M - AGLR Team)}
  \institution{Aix Marseille Univ, CNRS , I2M, Marseille, France}
  \streetaddress{163 Av. de Luminy}
  \city{Marseille}
  \postcode{13009}
  \country{France}                    
}
\email{valentin.maestracci@univ-amu.fr}

\author{Thomas Seiller}
\authornote{T. Seiller was partially supported by the \grantsponsor{Ile-de-France}{DIM RFSI}{Exploratory project} project \cohop, and the \grantsponsor{ANR}{ANR}{http://dx.doi.org/10.13039/501100001665} ANR-22-CE48-0003-01 project \dysco.}          
\orcid{0000-0001-6313-0898}             

\affiliation{
  \department[0]{LIPN -- UMR 7030 CNRS \& Sorbonne Paris Nord University}              
  \institution{CNRS}            
  \streetaddress{99, avenue Jean-Baptiste Clément}
  \city{Villetaneuse}
  \postcode{93430}
  \country{France}                    
}
\email{thomas.seiller@cnrs.fr}

\begin{document}

\maketitle

\section{Context and intuition}

Soon after the introduction of linear logic \cite{ll}, Girard proposed a research program \cite{towards} aiming at providing a mathematical representation of cut-elimination, or equivalently (through the proofs-as-programs correspondence) of program execution. This program, named geometry of interaction, quickly lead to the definition of several models \cite{multiplicatives,goi1,goi2} which in turn lead to the development of game semantics \cite{hylandong,AJM}. 
In early models, this mathematical operation was obtained through the so-called \emph{execution formula}, which was identified by Joyal, Street and Verity as an exemple of \emph{categorical trace} \cite{tracedmonoidal}. This work lead researchers to provide a categorical account of geometry of interaction based on traced monoidal categories \cite{haghverdi2000categorical}.

These models of geometry of interaction were studied from the point of view of providing a model of programs and their execution. 
However, a key aspect of the construction, which took more importance in later models, is that a model of (fragments of) linear logic could be defined on top of this dynamic representation of programs by realisability techniques. These techniques are on ideas similar to the definition of coherence spaces by means of an orthogonality relation \cite{doubleglueing,qcs} or the definition of realisability models over the lambda-calculus \cite{Riba}. As such, one would expect that they would fit the categorical framework of double gluing introduced by Hyland and Schalk \cite{doubleglueing}. However, to our knowledge, no geometry of interaction models have been shown yet to be an instance of double glueing.

As part of the geometry of interaction program, Interaction Graphs models were introduced by the second author in a series of papers \cite{seiller-goim,seiller-phd,seiller-goiadd,seiller-goig,seiller-goif,seiller-goie}. It provides a combinatorial approach to Girard's program.
One major conceptual contribution of Interaction Graphs was to shed light on a geometric identity underlying all previous geometry of interaction models introduced by Girard \cite{multiplicatives,goi1,goi2,goi3,feedback,goi5}. Indeed, all these models are recovered as instances of the \ig model for a specific choice of parameters. 
The underlying geometric identity, called the \emph{trefoil property} \cite{seiller-goiadd}, relates paths and cycles in the graph. As such, it generalises the usual "adjunction" in \goi models: the property that ensures monoidal closure of the induced category. The trefoil property turned out to be quite useful: beyond ensuring the monoidal closure, it can be exploited to define a model of additive connectives\footnote{It was a standard issue of geometry of interaction: since execution is defined locally, the cut-elimination steps between additive connectives is not represented "on the nose". Using the trefoil property, it can nevertheless be shown that these steps are represented up to behavioural equivalence \cite{seiller-goiadd}.}. But while the adjunction could be related to a categorical property, the trefoil property seemed to elude such interpretation. 

\paragraph{Contributions}


In this paper, we develop an idea mentioned in Seiller's PhD thesis \cite{seiller-phd}. Namely, that the two ingredients needed to define Interaction Graphs models -- associativity of execution and the trefoil property -- are the low dimensional projections of a single higher-dimensional associativity. We make this relation precise by considering categories of cobordisms. We recall that $\cob[n]$ is the category whose objects are $(n-1)$-dimensional manifolds and morphisms from $A$ to $B$ are $n$-dimensional manifolds with boundaries $A\disjun B$. Given such a cobordism $\cobord{M}$, one can associate a bipartite graph on $\pi_0(A)\disjun \pi_0(B)$ whose edges are paths -- up to homotopy -- between connected components of $A$ and connected components of $B$. One may also consider the set of cycles in $\cobord{M}$, i.e. the fundamental groupoid $\Pi_1(\cobord{M})$. The intuition is that the associativity of execution and the trefoil property are consequences of the associativity of composition in the category $\cob[n]$, through the two functors thus defined.

We show this intuition to be correct when working with low-dimensional cobordisms (namely the category $\cob[0]$). This formally presents (a submodel of) Interaction Graphs models as obtained through a double-gluing construction. We explain the difficulties arising in extending the analysis to higher-dimensional cobordisms, and propose a solution.

\section{Introduction}

\begin{definition}
The category $\cob[n]$ is defined as follows:
Objects are smooth manifolds of dimension $n$. The set $\homset[{\cob[n]}]{A}{B}$ of morphisms from $A$ to $B$ is the set of smooth manifolds $\cobord{M}$ of dimension $n + 1$ whose boundary $\partial\cobord{M}$ is equal to $A \disjun B$.

Composition in the category is given by gluing cobordisms along their shared boundaries. Formally, given $\cobord{M} \in \homset[{\cob[n]}]{A}{B}$ and $\cobord{N} \in \homset[{\cob[n]}]{B}{C}$, the cobordism $\cobord{M;N}$ is defined as the smooth manifold $\quotient{M \disjun N}{\sim}$ where $b_\cobord{M} \sim b_\cobord{N}$ for all $b \in B$.
\end{definition}

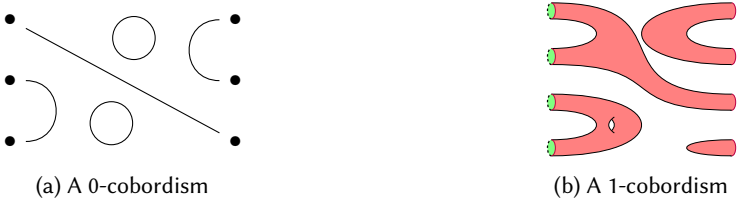
\begin{figure}
\begin{subfigure}[t]{0.49\textwidth}
\centering
\begin{tikzpicture}[x=1.5cm,y=0.8cm]
\node (a1) at (0,2) {$\bullet$};
\node (a2) at (0,1) {$\bullet$};
\node (a3) at (0,0) {$\bullet$};
\node (b1) at (2,2) {$\bullet$};
\node (b2) at (2,1) {$\bullet$};
\node (b3) at (2,0) {$\bullet$};
\node[draw,circle,inner sep=0.2cm] (c1) at (1.1,1.7) {~};
\node[draw,circle,inner sep=0.2cm] (c2) at (0.9,0.3) {~};
\draw[-] (a1) -- (b3) {};
\draw[-] (a2) .. controls (0.5,1) and (0.5,0) .. (a3) {};
\draw[-] (b1) .. controls (1.5,2) and (1.5,1) .. (b2) {};
\end{tikzpicture}
\subcaption{A 0-cobordism}\label{exemple-cob0}
\end{subfigure}
\begin{subfigure}[t]{0.49\textwidth}
\centering
\scalebox{.3}{
\begin{tikzpicture}[
  tqft,
  every outgoing boundary component/.style={fill=blue!50},
  outgoing boundary component 3/.style={fill=none,draw=red},
  every incoming boundary component/.style={fill=green!50},
  every lower boundary component/.style={draw,ultra thick, dashed},
  every upper boundary component/.style={draw,purple},
  cobordism/.style={fill=red!50},
  cobordism edge/.style={draw},
  view from=incoming,
  cobordism height=8cm,
]
\begin{scope}[every node/.style={rotate=90}]
\pic[name=a,
  tqft,
  incoming boundary components=2,
  genus=1,
  outgoing boundary components=0,
  ];
\pic[name=b,
  tqft,
  incoming boundary components=2,
  skip incoming boundary components={1},
  outgoing boundary components=1,
  offset=-1,
  at=(a-incoming boundary 2),
  anchor={(0,0)},
];
\pic[name=c,
  tqft,
  incoming boundary components=0,
  outgoing boundary components=2,
  at=(b-incoming boundary 1)
];
\pic[name=d,
  tqft,
  incoming boundary components=0,
  outgoing boundary components=1,
  at=(a-incoming boundary 1)
];
\end{scope}
\end{tikzpicture}
}
\subcaption{A 1-cobordism}
\end{subfigure}
\caption{Examples of cobordisms}\label{fig:cobordisms}
\end{figure}

We will be particularly interested in the category $\cob[0]$, whose objects are points and morphisms are segments and circles (see \autoref{exemple-cob0}). The point of this paper is to formally relate $\cob[0]$ to Seiller's Interactions graphs models. We therefore start by recalling basic constructions of the latter.

\begin{definition}
A directed graph $G$ is a tuple $(V^G,E^G,s^G,t^G)$, where $V^G$ is a finite set of vertices, $E^G$ is the set of edges, and $s^G, t^G$ -- the source and target maps -- are functions from $E^G$ to $V^G$. 
\end{definition}

\begin{definition}
A path in a graph $G$ is a sequence of vertices $e_1e_2\dots e_n$ such that for all $i\in \{1,\dots,n-1\}$, $s^G(e_{i+1})=t^G(e_i)$. The source $s^G(\pi)$ (resp. the target $t^G(\pi)$) of the path $\pi$ is defined as $s^G(e_1)$ (resp. $t^{G}(e_mn)$). A cycle in a graph $G$ is a path $\pi=e_1 e_2\dots e_n$ such that $s^G(\pi)=t^G(\pi)$.
\end{definition}


\begin{definition}
An alternating path between two graphs $G,H$ is a sequence of edges $e_1e_2\dots e_n$ such that for all $i\in\{1,\dots n-1\}$, $e_i \in E^G$ if and only if $e_{i+1}\in E^H$. The set of alternating paths between $G$ and $H$ will be denoted $\altpaths{G,H}$.

An alternating cycle is an alternating path such that $e_1\in E^G$ if and only if $e_n\in E^H$. A cycle is \emph{prime} if it not of the form $\rho^k$ for $k>1$, i.e. it is not the concatenation of several copies of the same cycle. The set of alternating prime cycles between $G$ and $H$ is denoted $\altprimecycles{G,H}$.
\end{definition}

\begin{definition}
The \emph{execution} of two graphs $G,H$ is the graph $G\plug H$ such that $V^{G\plug H}=V^G\symmdiff V^H$ (symmetric difference) and whose edges are the alternating paths of source and target in $V^G\symmdiff V^H$.
Alternatively, $G\plug H$ is the graph of finite maximal alternating paths between $G$ and $H$.
\end{definition}

One can check that execution endows the category of graphs bipartite graphs (where composition is given by computing paths of length 2) with a categorical trace. From this we can define a category of interaction graphs as an instance of the Int construction \cite{tracedmonoidal}.

\begin{definition} Objects of $\igrph$ are finite sets. A morphism $F: A \to B$ is a graph on $A + B$, and composition is defined by the execution formula.
\end{definition}

However we note that the category $\igrph$ is unsufficient to constrcut a model of multiplicative linear logic. One needs "to extend it" to interpret proofs as a pair (called a \emph{project}) $(a,A)$ of a real number $a$ -- the wager -- and a graph $A$. Then the notion of execution is extended from graphs to projects as follows: $(a,A)\plug(b,B)=(a+b+\meas{A,B},A\plug B)$, where $\meas{A,B}$ is a parametrized measure of prime cycles, possibly using their weights. Since we consider here unweighted graphs, this measure can only count prime cycles, i.e. $\meas{A,B}=\card{\altprimecycles{A,B}}$.
From these, one can define a model by realisability techniques. We refer the interested reader to the original papers for more details \cite{seiller-goim,seiller-goiadd}.

Our goal is now to show that the following two notable properties, which are essential in constructing \ig models, are the image of a higher-dimensional associativity (namely the associativity of composition in $\cob[n]$). Given three graphs $F,G,H$ such that $V^F\cap V^G\cap V^H=\emptyset$:
\begin{itemize}[nolistsep,noitemsep]
\item \textbf{Associativity of execution:}
\( (F\plug G)\plug H \cong F\plug (G\plug H);\)
\item \textbf{Trefoil property:} 
\( \altprimecycles{F,G\plug H}\disjun\altprimecycles{G,H} \simeq \altprimecycles{H,F\plug G}\disjun\altprimecycles{F,G}.\)
\end{itemize}

\section{$\cob[0]$ : A simple, working case}

Looking back at our example of $\cob[0]$, a naive approach would be to take as functor the fundamental groupoid. But this does not work, for this functor would not map the identity to the identity. This can be corrected by excluding self-loops. While this constraint seems ad-hoc, we note the proposed solution for higher-dimensional cobordisms will provide an alternative, more satisfying, solution to this problem.

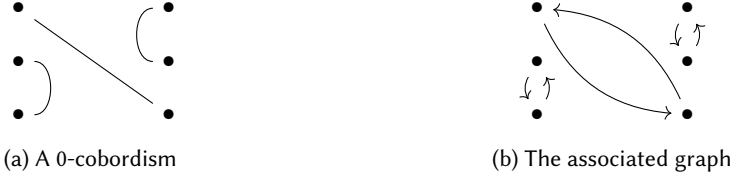
\begin{figure}
\begin{subfigure}[t]{0.49\textwidth}
\centering
\begin{tikzpicture}[x=2cm,y=1.4cm,scale=0.5]
\node (a1) at (0,2) {$\bullet$};
\node (a2) at (0,1) {$\bullet$};
\node (a3) at (0,0) {$\bullet$};
\node (b1) at (2,2) {$\bullet$};
\node (b2) at (2,1) {$\bullet$};
\node (b3) at (2,0) {$\bullet$};
\draw[-] (a1) -- (b3) {};
\draw[-] (a2) .. controls (0.5,1) and (0.5,0) .. (a3) {};
\draw[-] (b1) .. controls (1.5,2) and (1.5,1) .. (b2) {};
\end{tikzpicture}
\subcaption{A 0-cobordism \label{exemple-cob0-2}}
\end{subfigure}
\begin{subfigure}[t]{0.49\textwidth}
\centering
\begin{tikzpicture}[x=2cm,y=1.4cm,scale=0.5]
\node (a1) at (0,2) {$\bullet$};
\node (a2) at (0,1) {$\bullet$};
\node (a3) at (0,0) {$\bullet$};
\node (b1) at (2,2) {$\bullet$};
\node (b2) at (2,1) {$\bullet$};
\node (b3) at (2,0) {$\bullet$};

\node (mida1b3) at (1,1){};

\draw[->] (a1) to[bend right] (b3);
\draw[->] (b3) to[bend right] (a1);

\draw[->] (a2) to[bend right] (a3);
\draw[->] (a3) to[bend right] (a2);

\draw[->] (b1) to[bend right] (b2);
\draw[->] (b2) to[bend right] (b1);
\end{tikzpicture}
\subcaption{The associated graph\label{exemple-F-cob0}}
\end{subfigure}
\caption{The path functor $\F$}\label{fig:functor}
\end{figure}

%
%
%
%
%
%
%
%

\begin{definition}
Let $\cobord{M}$ be a cobordism in $\homset[{\cob[1]}]{A}{B}$. We define its \emph{fundamental graph} $\Gamma_1(\cobord{M})$: 
\begin{itemize}[noitemsep,nolistsep]
\item $V^{\Gamma_1(\cobord{M})}=A \disjun B$,
\item $E^{\Gamma_1(\cobord{M})}=\set{[p] \mid p: [0;1] \to \cobord{M}, p(0) \neq p(1)}$,
\item $s^{\Gamma_1(\cobord{M})}([p])=p(0)$, and $t^{\Gamma_1(\cobord{M})}([p])=p(1)$.
\end{itemize}
We also consider $\F : \cob[0] \to \igrph$ acting as the identity on objects, and as $\Gamma_1$ on morphisms.
\end{definition}

Functoriality of $\F$ is a consequence of the following result, whose proof is essentially the first part of the proof of the Van-Kampen theorem.

\begin{lemma}
A path $p : \I \to C_1 ; C_2$ has a unique (up to homotopy) decomposition $p_1, \cdots p_n$ as alternating paths, ie with $p_{i} \subseteq C_{\delta(i)}$, $\delta(i) \neq \delta(i+1)$ for a certain $\delta : [1;n] \to \{1,2\}$    
\end{lemma}


\begin{theorem}
The mapping $\F : \cob[0] \to \igrph$ is a functor.
\end{theorem}

However, this functor is not faithful as circles, the boundaryless components of the cobordism, are entirely forgotten. For instance, the cobodisms shown in \autoref{exemple-cob0} and \autoref{exemple-cob0-2} are both mapped to the graph shown in \autoref{exemple-F-cob0}.

To extend it to a faithful functor, one can consider pairs of a graph and an integer counting the numbers of such circles. This is exactly the role of \emph{wagers} in interaction graphs models \cite{seiller-phd}. 
From this observation, we define an extended functor
$\Fb : \cob[0] \to \project$ acting as the identity on objects and mapping a cobordism $\cobord{M}$ to $(\F(\cobord{M}), n_\cobord{M})$ on morphisms, with $n_\cobord{M}$ the number of loops in $\cobord{M}$. Note that $n_\cobord{M}$ is characterised by the fundamental group of $\cobord{M}$: $\pi_1(\cobord{M})=\mathbb{Z}^{n_\cobord{M}}$. 

\begin{theorem}
    The extended functor $\Fb$ is faithful.
\end{theorem}

This result has two consequences. Firstly, the bi-orthogonality construction that was used in \cite{seiller-goim} was -- at least on the subcategory $\Fb(\cob[0])$ -- a case of Hyland and Schalk \emph{tight} double-glueing w.r.t. a \emph{focused orthogonality} \cite{doubleglueing}.
    Secondly, both the trefoil property and the associativity of execution are but reflections of the associativity of composition in $\cob[0]$ in lower dimensions.

\section{Higher dimensions}

\subsection{Problems arising}

The case of $\cob[0]$, while interesting, is extremely limited. Since we are studying paths on surfaces, the obtained graphs will always be symmetric, but only pairings are obtained from morphisms in $\cob[0]$, i.e. vertices are of degree exactly one. We can expect to obtain more graphs by considering higher cobordisms categories; in fact this is already different in $\cob[1]$.
%
%
Can we generalize the previous section to $\cob[n]$, with $n\leqslant 1$? Multiple obstacles appear if one tries to adapt the proof that $\F$ is a functor.

First, one needs to define the vertices of the graph; where we had a point in $\cob[0]$, we now have an entire manifold. The simplest solution seems to take a vertex for each connected component of the boundary manifolds, together with a representative point that would serve as base point for the paths.

\begin{figure}
\begin{subfigure}[t]{0.99\textwidth}
\centering

\includegraphics[scale = 0.38]{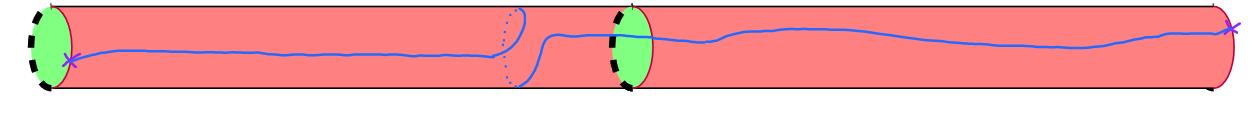}

\subcaption{A path}
\end{subfigure}
\begin{subfigure}[t]{0.99\textwidth}
\centering

\includegraphics[scale = 0.38]{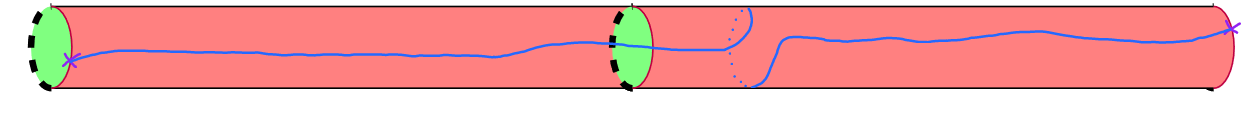}

\subcaption{An homotopy equivalent path}
\end{subfigure}
\caption{Counterexample to functoriality}\label{fig:functor_problem}
\end{figure}

Second, one would like to decompose a path in $\F(C_1;C_2)$ as a finite sequence of alternating paths of $\F(C_1)$ and $\F(C_2)$, but the situation is more complex. 
Here, this decomposition is not unique, as illustrated in \autoref{fig:functor_problem} where we shown two paths that are homotopy equivalent in the composition but can be decomposed in two different ways as a composition of two paths. Hence, considering path only up to homotopy doesn't give rise to a functor.

\subsection{Our proposal for a solution}

One way to solve this issue would be to associate to a cobordism the set of all paths -- not equivalence classes up to homotopy. This option is not viable, since if the circle was located on the border we wouldn't have a unique decomposition. We therefore chose to follow a second option, based on the introduction of a higher dimensional structure in the category of graphs that will allow to identify compositions of paths such as shown in \autoref{fig:functor_problem}. The intuition behind the formalism is that the fundamental group of the border of the cobordism acts on equivalence classes of paths up to homotopy both by pre-composition and post-composition. The set of paths up to homotopy can therefore be considered as a set endowed with a right and a left action -- much like a bi-module. The composition of paths should therefore be quotiented by an equivalence akin to the quotient performed in the definition of tensor product of bi-modules: $[p]\cdot a \sim a\cdot [p]$.

We therefore define a categorical structure in which each object $A$ is associated with a group $G_A$, and the set of morphisms $\homset[]{A}{B}$ is endowed with a left action by $G(A)$ and a right action by $G(B)$. Composition is then defined up to the following identity: the composition of $f\in \homset[]{A}{B}$ and $g\in \homset[]{B}{C}$ is the equivalence class of $f;g$ (to avoid left/right confusion, we note composition sequentially) w.r.t. the identification of $f \cdot b; g$ with $f ; b\cdot g$.

\begin{definition}
A bimodular graph is given as a tuple $(V^G,E^G,s^G,t^G,\gamma^G,\lambda^G,\rho^G)$, where $(V^G,E^G,s^G,t^G)$ is a directed graph, and:
\begin{itemize}[noitemsep,nolistsep]
\item $\gamma^G: V^G\to\mathbf{Group}$ associate to each vertex a group;
\item $\lambda^G$ maps pairs of vertices to left actions on edges between those, i.e. $\forall v,v'\in V^G$, $\lambda^G(v,v')$ defines a left action of $\gamma^G(v)$ on $E^G(v,v')$;
\item $\rho^G$ maps pairs of vertices to right actions on edges between those.
\end{itemize}
\end{definition}

One can define a standard notion of composition of modular graphs in a similar way as for graphs, as soon as the maps $\gamma$ coincide on the common vertices: we define the composition as the paths of length 2 modulo the identification informally explained above. Formally if $e,e'$ are edges in $E^F(v,v')$ and $E^{G}(v',v'')$ respectively, we identify the path $ee'$ with $\rho(b)(e)\lambda(b^{-1})(e')$ for all $b\in\gamma^G(v')$.

The category of modular graphs defined in this way can be shown to be traced monoidal, defining execution in the same way as before. The category obtained by the Int construction then generalises the category of interaction graphs in a way that allows for identifying some compositions of paths, hence avoiding the issue of non-unicity of decomposition shown in \autoref{fig:functor_problem}. It therefore provides a good candidate to extend the mapping defined in the previous section from $\cob[0]$ to $\cob[1]$, and possibly to higher dimensional cobordisms.

\section{Future directions}

This work opens up several directions that we would like to explore.

Firstly, cobordisms -- and glueing of cobordisms -- are a particular case of the categorical notion of cospan. We can envision to generalize our approach to more general topological spaces (e.g. simplicial sets) using cospans.

Secondly, this work only captures the case of symmetric graphs. It may be possible to extend the techniques to obtain general directed graphs by considering directed spaces, and using methods from directed algebraic topology.

Finally, one interesting aspect of cobordisms categories is that $\cob[n]$ is a the category of morphisms between identities in $\cob[n-1]$. As a consequence, consideration of the family of models obtained from $\cob[n]$ for all $n$ could be of interests to approach the question of linear dependent types.

    


\bibliographystyle{abbrv}
\bibliography{bibliography}

\end{document}